\documentclass[10pt,a4paper]{article}

\usepackage{graphicx, natbib, amssymb}

\usepackage[centertags]{amsmath}

\numberwithin{equation}{section}

\normalsize \textwidth 454pt \textheight 690pt \baselineskip 14.5pt \columnsep 12pt \columnwidth
230pt \topskip 12pt \evensidemargin 26pt \oddsidemargin 9pt \topmargin -5 mm

\begin{document}

\title{Generalized information-entropy measures and Fisher information}

\author{Marco Masi\footnote{Corresponding author: marco\!\_masi2@tin.it, fax: +39 02 6596997} \\ \small via Varese 16, 20121, Milan, Italy.}

\maketitle

\begin{abstract}
We show how Fisher's information already known particular character as the fundamental information
geometric object which plays the role of a metric tensor for a statistical differential manifold,
can be derived in a relatively easy manner through the direct application of a generalized
logarithm and exponential formalism to generalized information-entropy measures. We shall first
shortly describe how the generalization of information-entropy measures naturally comes into being
if this formalism is employed and recall how the relation between all the information measures is
best understood when described in terms of a particular logarithmic Kolmogorov-Nagumo average.
Subsequently, extending Kullback-Leibler's relative entropy to all these measures defined on a
manifold of parametrized probability density functions, we obtain the metric which turns out to be
the Fisher information matrix elements times a real multiplicative deformation parameter. The
metrics independence from the non-extensive character of the system, and its proportionality to
the rate of change of the multiplicity under a variation of the statistical probability parameter
space, emerges naturally in the frame of this representation.
\end{abstract}



\newpage

\section{Introduction}

In a previous paper \cite{Masi}, using a formalism based on Kolmogorov-Nagumo means and
generalized logarithms and exponentials, we wrote down the set of entropy functionals, from
Boltzamann-Gibbs entropy through R\'{e}nyi and Tsallis, up to Sharma-Mittal \cite{Sharma} and a
new entropy measure, we called the "supra-extensive entropy", so that the increasing
generalization of entropy measures from arithmetic to non-arithmetic means, and from extensive to
non-extensive systems became particularly compact and visible in its hierarchical structure.
Sharma-Mittal measure was already developed in 1975 but has been investigated in generalized
thermostatistics only recently by Frank, Daffertshofer and Naudts (\cite{Daffertshofer},
\cite{Frank} \cite{Naudts}). We showed that Sharma-Mittal's measure is however only one of two
possible extensions that unify R\'{e}nyi and Tsallis entropy in a coherent picture and described
how it comes naturally into being together with another "supra-extensive" measure if the formalism
of generalized logarithm and exponential functions is used. Moreover, we could see how the
relation between these information measures is best understood when described in terms of a
logarithmic Kolmogorov-Nagumo average.

In this paper we shall further investigate in particular the power of the deformed
logarithm-exponential formalism with regards to the relationship of generalized entropy measures
and Fisher information.

Fisher information was originally conceived in the 1920s \cite{Fisher1}, many years before
Shannon's notion of entropy, as a tool of statistical inference in parameter estimation theory. It
must be emphasized that Fisher's functional is an information, but not an entropy measure. There
is nevertheless a strong connection between Fisher information and entropy. This relationship has
been outlined in many occasions since Rao \cite{Rao}, already in 1945, laid the foundations of
statistical differential geometry, called also \textit{information geometry}  (for a more recent
review of the subject see e.g. Amari \& Nagaoka \cite{amari}). Rao outlined how a statistical
model can be described by a statistical differential manifold which can be considered as a
Riemannian manifold of parametrized probability distributions (PD) or probability density
functions (PDF) with the metric tensor given by the Fisher information matrix (FIM). The FIM
determines a Riemannian information metric on this parameter space, and is therefore called also
the \textit{Fisher metric}. This has been the subject of renewed interest more recently also in
other branches of information theory, in applications of image processing, econometrics and
received some attention in theoretical physics, especially in regards to its, still not entirely
understood, role in quantum mechanics and perhaps also quantum gravity (see e.g. B.R. Frieden's
work \cite{Frieden} which tries to derive the laws of physics from a Fisherian point of view, or
R. Carol's review \cite{Carrol} of some other similar attempts and references therein).

Less has been done to highlight the links between Fisher information and generalized measures and
non-extensive statistics. Some attempts in this direction were made for instance by F. Pennini and
A. Plastino \cite{Pennini}, M. Portesi, F. Pennini and A. Pennini \cite{Pennini2}, S. Abe
\cite{Abeprova}, J. Naudts (\cite{Naudtsprova1}, \cite{Naudtsprova2}), P. Jizba \cite{Jizbaprova},
just to mention some examples. However, we feel that a clear exposition is lacking about the place
that the Fisher information measure has in the frame of a generalized statistics. The aim of this
paper is to highlight in a synthetic way the relationship that exists between Fisher information
and the two-parametric generalized entropy measures here mentioned (R\'{e}nyi, Tsallis, Sharma-
Mittal and the supra-extensive measure which expands further the picture as a consequence of the
q-deformed formalism) in the sense that diagram of page 14 illustrates, what role the two
parameters play in evaluating the Fisher information matrix, and how it can be retrieved using a
deformed exponential formalism. We will focus our attention on how precisely Fisher information
(except a real multiplicative factor) emerges naturally as a universal statistical metric tensor
for every generalized information-entropy measure defined on a manifold of PDFs (i.e. for a
continuous version of the above mentioned entropies), and to obtain in a relatively simple manner
this result using a representation based on generalized logarithm and exponential functions within
the frame of a KN formalism.

It should also be mentioned that R\'{e}nyi entropy is not Lesche-stable \cite{Lesche}, isn't
convex and does not possess the property of finite entropy production. Therefore any extension of
R\'{e}nyi's entropy, cannot in general possess these properties either. There is some controversy
if this is supposed to have its thermodynamical implications, or not. However, the theoretical
framework we are going to construct here has to be intended in a more general context, it can
still have its meaning and applications in information theory, cybernetics or other fields not
necessarily restricted to a generalized thermostatistics. It is with this point of view in mind
that we will proceed.

\section{The generalized information-entropy measures}

Just to make this paper selfcontained let us briefly sum up the aspects of a generalized
information-entropy measure theory which will be relevant to the understanding of the next
sections.

\subsection{The Boltzmann-Gibbs entropy and Shannon's information measure}

As is well known the Boltzmann-Gibbs (BG) entropy reads\footnote{Here we begin to introduce a more
general symbolism according to which every type of information measure is labeled with
$S_{name}(\{P\},\{q\})$  or $\mathcal{S}_{name}(\mathcal{\{P\}},\{q\})$, where $P$ or
$\mathcal{P}$ stands for the family $\{p_{i}\}$ of PDs or PDFs and $S$ or $\mathcal{S}$ for the
discrete and continuous cases respectively, while $q$ is a scalar or vector parameter which
meaning will become clear in the following sections.} $$S_{BG}(P)= - k \, \sum_{i} p_{i} \log
p_{i} \, ,$$ with $p_{i}$ the probability of the system to be in the i-th microstate, $k$ the
Boltzmann constant. BG entropy becomes the celebrated Shannon information measure \cite{Shannon}
if $k=1$ (as we will do from now on) and uses the immaterial base $b$ for the logarithm function
(we will maintain the natural logarithm $b=e$)\begin{equation}S_{S}(P) = - \sum_{i} p_{i} \log_{b}
p_{i} = \sum_{i} p_{i} \log_{b} \left( \frac{1}{p_{i}} \right) \equiv \sum_{i} p_{i} \log \left(
\frac{1}{p_{i}} \right) \, .\label{shannon}
\end{equation} BG and Shannon's measures are additive, i.e. given two systems, described by two
PDs $A$ and $B$, we have $$S_{S}(A \cap B) = S_{S}(A) + S_{S}(B|A) \, ,$$ with $S_{S}(B|A)$ the
conditional entropy. These systems are called \textit{extensive systems}. This is the case where
the total entropy behaves as the sum of the entropies of its parts and applies to standard
statistical mechanics. The additive property is reflected in the logarithm function.

\subsection{Tsallis' entropy}

Nature is however not always a place where additivity is preserved. This is the case of nonlinear
complex systems, in fractal- or multifractal-like and self-organized critical systems, or where
long range forces are at work (e.g. in star clusters or in systems with long range microscopic
memory), etc. These \textit{non-extensive systems} have been investigated especially in the last
two decades \cite{Tsallis2}.

Tsallis generalized Shannon's entropy to non-extensive systems as \cite{Tsallis} \begin{equation}
S_{T}(P,q) = \frac{\sum_{i}p_{i}^{q}-1}{1-q} \, = \, \frac{1}{q-1} \sum_{i}p_{i} \, (1-
p_{i}^{q-1}) \, ,\label{tsallis} \end{equation} with $q$ a real parameter. This is now widely
known as \textit{Tsallis entropy}. According to a current school of thought at least some
non-extensive systems can be described by scaled power law probability functions as $p_{i}^{q}$,
so called \textit{q-probabilities}. For $q\rightarrow1$ it reduces to Shannon's measure. Tsallis
entropy extends additivity to \textit{pseudo-additivity}
\begin{equation}S_{T}(A \cap B) = S_{T}(A) + S_{T}(B|A) + (1-q)S_{T}(A) S_{T}(B|A) \, .\label{tadditive} \end{equation}

In order to describe Tsallis sets the \textit{generalized q-logarithm function}\begin{equation}
\log_{q} x = \frac{x^{1-q}-1}{1-q} \, , \label{q-log} \end{equation} turns out to be particularly
useful. In a similar way, its inverse, the \textit{generalized q-exponential function} is
\begin{equation} e_{q}^{x} = [1+(1-q)x]^{\frac{1}{1-q}} \, .\label{q-exp} \end{equation} The
classical Napier's logarithm and its inverse function is recovered for $q=1$. The importance of
the q-logarithm in this context is realized if we understand that it satisfies precisely a
pseudo-additive law $$\log_{q} xy = \log_{q} x + \log_{q} y + (1-q) (\log_{q} x) (\log_{q} y) \,
.$$

Exploiting this generalized logarithm and exponential formalism Tsallis entropy \ref{tsallis} can
be rewritten as \begin{equation} S_{T}(P,q) = - \sum_{i}p_{i}^{q} \log_{q}p_{i} = \sum_{i} p_{i}
\log_{q}\left(\frac{1}{p_{i}}\right) \, , \label{q-shannon}\end{equation} which is sometimes also
referred to as the \textit{q-deformed Shannon entropy}.

Note that $\log_{q}x^{\alpha} \neq \alpha \log_{q}x$ when $q\neq 1$. This is the reason why, if
one thinks in terms of averages, it is more meaningful to write entropy measures with the inverse
of the PD, as in the r.h.s. of \ref{q-shannon}, and why we will prefer this formal representation.

\subsection{R\'{e}nyi's entropy}

By looking at the structure of the r.h.s. of \ref{shannon} and \ref{q-shannon} one can define an
information measure as an average of the \textit{elementary information gains}
\begin{equation}I_{i} \equiv I_{i}\left(\frac{1}{p_{i}}\right) = \log_{q}\left(\frac{1}{p_{i}}\right)
\label{s2} \end{equation} associated to the i-th event of probability $p_{i}$
\begin{equation}S_{S}(P) = \left< \log \left(\frac{1}{p_{i}}\right)
\right>_{lin}\label{shannon2}\end{equation} and
\begin{equation}S_{T}(P) = \left< \log_{q} \left(\frac{1}{p_{i}}\right)
\right>_{lin}\label{tsallis2}\end{equation} where, what is common to both, is the underlying
\textit{arithmetic-}, or \textit{linear mean} $I = \sum_{i} p_{i} I_{i} $.

However, A.N. Kolmogorov and M. Nagumo (\cite{Kolmogorov2}, \cite{Nagumo}) showed, already in 1930
but independently from each others that, if we accept Kolomogorov's axioms as the foundation of
probability theory, then the notion of average can acquire a more general meaning as what is
called a \textit{quasi-arithmetic} or \textit{quasi-linear mean}, and can be defined as
\begin{equation} S = f^{-1} \left( \sum_{i} p_{i} \, f(I_{i}) \right) \, , \label{infomean}
\end{equation} with $f$ a strictly monotonic continuous function, called the
\textit{Kolmogorov-Nagumo function} (KN-function). R\'{e}nyi instead showed \cite{Renyi} that, if
additivity is imposed on information measures, then the whole set of KN-functions must reduce to
only two possible cases. The first is of course the linear mean associated with the KN-function
$$f(x) = x \, , $$ while the other possibility is the \textit{exponential mean} represented by the KN-function \begin{equation}f(x) = c_{1} \,
b^{(1-q)x} + c_{2} \, ; \hspace{10mm} q \in \mathbb{R}\label{quasilin} \end{equation} with $c_{1}$
and $c_{2}$ two arbitrary constants.

\textit{R\'{e}nyi's information-entropy measure} is per definition a measure where the single
information gains are averaged exponentially, and writes \begin{equation} S_{R}(P,q) =
\frac{1}{1-q} \log_{b} \sum_{i} p_{i}^{q} \equiv \frac{1}{1-q} \log \sum_{i} p_{i}^{q} \, ,
\label{renyi} \end{equation} with $b$ the logarithm base (still we will always assume $b=e$). When
$q\rightarrow1$ R\'{e}nyi's boils down to Shannon entropy.

In fact, if we choose in \ref{quasilin}, $c_{1}=\frac{1}{1-q}=-c_{2}$, then because of
\ref{q-log}, it becomes \begin{equation}f(x) = log_{q} \, e^{x} \, ,\label{renyifun}
\end{equation} which inserted in \ref{infomean} with $$I_{i}=\log \left(\frac{1}{p_{i}}\right) \, ,$$
shows that \ref{renyi} is equivalent to $$S_{R}(P,q) = \left< \log \left( \frac{1}{p_{i}} \right)
\right>_{\!\!\mathrm{exp}} \, ,$$ where $\left<\cdot\right>_{exp}$ stands for an average defined
by KN-function \ref{renyifun}. Compare this with \ref{shannon2} and \ref{tsallis2}.

\subsection{The Sharma-Mittal and \textit{Supra-extensive} entropy}

The next step in the generalization process consists in finding a measure which is non-extensive
and non-additive but contains Tsallis' and R\'{e}nyi's entropies as special cases. One possible
way to obtain this goes through an extension of the KN-mean. This leads to what is known as the
Sharma-Mittal entropy (SM) \cite{Sharma}. However it is only by exploiting the generalized
logarithm and exponential representation one retrieves in a compact and fast manner both SM
entropy measure and what we used to call the "supra-extensive" (SE) entropy. By using the
q-deformed logarithm and exponential formalism one could easily arrive at a further generalization
of R\'{e}nyi and Tsallis entropies.

The starting point is the relationship between Tsallis and R\'{e}nyi entropies $$S_{R}(P,q)
=\frac{1}{1-q} \log \left[ 1+ (1-q) \, S_{T}(P,q) \right] \, . $$

From \ref{q-log} and \ref{q-exp}, we see that this is equivalent to
\begin{equation}S_{R}(P,q) = \log e_{q}^{S_{T}(P,\,q)} \, ,\label{ireqt}
\end{equation} and therefore \begin{equation}S_{T}(P,q) = \log_{q}e^{S_{R}(P,\,q)} \, . \label{itqir}
\end{equation}

\ref{ireqt} and \ref{itqir} suggest immediately two further generalization:
\begin{equation}S_{SM}(P,\{q,r\})=\log_{r} e_{q}^{S_{T}(P,\,q)} = \frac{1}{1-r} \left[ \left(
\sum_{i} p_{i}^{\,q} \right) ^{\frac{1-r}{1-q}} - 1 \right] \, , \label{eqit}
\end{equation} and \begin{equation}\hspace{9mm} S_{SE}(P,\{q,r\})= \log_{q}e_{r}^{S_{R}(P,\,q)} =
\frac{\left[ 1 + \frac{(1-r)}{(1-q)} \, \log \sum_{i}p_{i}^{q}\right]^{\frac{1-q}{1-r}}-1}{1-q}\,
,\label{qrir}\end{equation}

with $r$ another real parameter.

\ref{eqit} is SM's pseudo-additive measure, while \ref{qrir} is a new type of entropy measure we
called "supra-extensive" because it generalizes to a measure which is neither additive nor
pseudo-additive. We could see \cite{Masi} how the decisive difference between these two
information-entropies is that SM's measure can be obtained also through the KN-mean as a two
parameter extension of \ref{renyifun} (with $f(x) = log_{q} \, e_{r}^{x}$ on $I_{i}=\log_{r}
\left(\frac{1}{p_{i}}\right)$), while the SE measure does not have such kind of generalization. It
can also be shown that for two systems $A$ and $B$ for Sharma-Mittal entropy (instead of
\ref{tadditive}) one has $$S_{SM}(A \cap B) = S_{SM}(A) + S_{SM}(B|A) + (1-r)S_{SM}(A) S_{SM}(B|A)
\, .$$ This indicates that it is the magnitude of parameter $r$ which stands for the degree of
non-extensivity, and $q$ stands for a PD deformation parameter. When $r \rightarrow q$ the
deformation parameter $q$ of the PD merges into the non-extensivity parameter $r$ (which is the
reason why in Tsallis entropy it is $q$ instead of $r$ that appears for the non-extensive
character of the system).

The supra-extensive entropy \ref{qrir} however emerges naturally as a symmetric counterpart of
\ref{eqit} when generalized logarithms and exponentials are used. Further mathematical-physical
investigations which will clarify the standpoint of the supra-extensive entropy, what kind of
statistics it expresses, if any, and its relationship with other measures, is of course desirable
and still necessary. Anyway, something can be already said. What we are going to do here is that
we can show how this new entropy also shares a common status in regards to Fisher information with
all the other measures too.

\subsection{The multiplicity}

To introduce ourselves to this, note first of all how we can rewrite the quantity
\begin{equation}\left(\sum_{i}p_{i}^{q}\right)^{\frac{1}{1-q}} = \left(\sum_{i}p_{i} \left(
\frac{1}{p_{i}} \right) ^{1-q}\right)^{\frac{1}{1-q}} = \left<
\left(\frac{1}{p_{i}}\right)^{1-q}\right>^{\frac{1}{1-q}}_{\!\mathrm{lin}} =
e_{q}^{\left<log_{q}\left(\frac{1}{p_{i}}\right)\right>_{\!\mathrm{lin}}}$$ $$ =
e_{q}^{S_{T}(P,q)}= \left<\frac{1}{p_{i}}\right>_{\!\!\log_{q}} \equiv \Omega(P,q) \,
,\label{eii}\end{equation} where we used what we call the \textit{logarithmic mean} $\left< \cdot
\, \right>_{log_{q}}$ defined by the KN-function $f(x)=log_{q}x$.

Then using \ref{eii}, equations \ref{ireqt} to \ref{qrir} can be rewritten as
\begin{equation}\hspace{-0mm} S_{T}(P,q)  = \log_{q} \left<\frac{1}{p_{i}}\right>_{\!\!\log_{q}} =
\log_{q} \Omega(P,q) \, ;\label{nonreltsallis}\end{equation}

\begin{equation}
\hspace{-0mm} S_{R}(P,q) = \log \left<\frac{1}{p_{i}}\right>_{\!\!\log_{q}} = \log \Omega(P,q)\,
;\label{nonrelrenyi}
\end{equation}

\begin{equation}\hspace{-0mm} S_{SM}(P,\{q,r\}) = \log_{r}
\left<\frac{1}{p_{i}}\right>_{\!\!\log_{q}} = \log_{r} \Omega(P,q)  \,
;\label{nonrelnaudts}\end{equation}

\begin{equation}\hspace{-0mm} S_{SE}(P,\{q,r\}) = \log_{q}
e_{r}^{\,\log \left<\frac{1}{p_{i}}\right>_{\!\log_{q}}} = \log_{q}e_{r}^{\,\log \Omega(P,\,q)} \,
. \label{nonrelmasi}\end{equation}

Rewriting things in the language of this representation and using the KN logarithmic mean one can
see more straightforwardly how Sharma-Mittal's entropy generalizes R\'{e}nyi's extensive entropy
to non-extensivity, and how the new measure does the same for non-extensivity generalizing it to a
'generalized non-extensivity', we called \textit{supra-extensivity}.

The quantity $$\Omega (P,q) = \left<\frac{1}{p_{i}}\right>_{\!\log_{q}} =
\left(\sum_{i}p_{i}^{\,q}\right)^{\frac{1}{1-q}} \, ,$$ is well known to have a physical
interpretation in statistical mechanics: the multiplicity of the system, i.e. the number of all
possible microstates compatible with its macroscopic state.

\section{Generalizing to relative entropy-information measures}

S. Kullback and R. A. Leibler \cite{Kuleibler} introduced the notion of \textit{relative entropy}.

Given a random variable $X$ with $\mathbf{x}$ a specific (scalar or vector) value for $X$ on a
continuous event space, consider continuous differentiable PDFs, $p(x,\theta)\in \mathcal{C}^{2}$,
with $\mathbf{\theta}$ a (scalar or vector) parameter. Let be $H_{1}$ the hypotheses that $X$ is
from the statistical population with PDF $p_{1}(\mathbf{x,\theta})$ and $H_{2}$ that with PDF
$p_{2}(\mathbf{x,\phi})$. Then it can be shown \cite{Kullback} that applying Bayes' theorem, $\log
\frac{p_{1}(\mathbf{x,\theta})}{p_{2}(\mathbf{x,\phi})}$ measures the difference between the
\textit{logarithm of the odds in favor of $H_{1}$ against $H_{2}$} before a measurement gave
$X=\mathbf{x}$. \textit{Kullback's relative entropy}, or our "mean capacity for discrimination" in
favor of $H_{1}$ against $H_{2}$, was originally defined as $$\mathcal{E}_{KL} =
\mathcal{E}_{KL}(\{p_{1},p_{2}\}) = \int_{\mathcal{S}_{sp}} \, p_{1}(\mathbf{x,\theta}) \log
\frac{p_{1}(\mathbf{x,\theta})}{p_{2}(\mathbf{x,\phi})}\, d^{n}\mathbf{\!x} \, ,$$ with
$\mathcal{S}_{sp}$ the entire sample space.

If $p_{2}(\mathbf{x,\phi})=1$ (we "discriminate" against certainty), the negative Shannon
information (in its continuous form) is recovered. The different signature is due to the fact that
Shannon's information, as all the measures we are dealing with here, account for the amount of
information we still need to gain complete knowledge, i.e. the uncertainty about the message. Let
us therefore call \textit{Kullback-Leibler relative information-entropy measure}, or simply
\textit{Kullback's measure} \begin{equation} \mathcal{S}_{KL} = \mathcal{S}_{KL}(\{p_{1},p_{2}\})
= \int_{\mathcal{S}_{sp}} \, p_{1}(\mathbf{x,\theta}) \log
\frac{p_{2}(\mathbf{x,\phi})}{p_{1}(\mathbf{x,\theta})}\, d^{n}\mathbf{\!x} \, .
\label{negkullback} \end{equation}

Relative entropies can be used to generalize all information measures either in their continuous
as in their discrete version. Let us start first with discrete PDs.

Given two families of PDs $P=\{P^{(1)};P^{(2)}\} = \{p_{i}^{(1)} \, ; \, p_{j}^{2}\}$,
$(i,j=(1,...,\Omega)$), Kullback's measure \ref{negkullback} takes the form
\begin{equation}S_{KL}(P) = \sum_{i}p_{i}^{(1)} \log \left(
\frac{p_{i}^{(2)}}{p_{i}^{(1)}} \right) = \left< \log
\left(\frac{p_{i}^{(2)}}{p_{i}^{(1)}}\right)\right>_{\!\mathrm{lin}} = \log \left<
\left(\frac{p_{i}^{(2)}}{p_{i}^{(1)}}\right)\right>_{log} \, .\label{disnegkullback}
\end{equation}

Then, in a more general context, we can extend \ref{s2} to \textit{elementary relative information
gains} as $$I_{i}=\log \left( \frac{p_{i}^{(2)}}{p_{i}^{(1)}}\right) \hspace{5mm} \mathrm{(for
\,\, extensive \,\, systems)} \, ,$$ or $$\hspace{10mm} I_{i}=\log_{s} \left(
\frac{p_{i}^{(2)}}{p_{i}^{(1)}}\right) \hspace{5mm} \mathrm{(for \,\, non-extensive \,\, systems)}
\, ,$$ in $$I = f^{-1} \left( \sum_{i} p_{i}^{(1)} \, f(I_{i}) \right) \, ,$$ with $s=q$ or $s=r$
for Tsallis' and SM's entropies respectively, that is we can rewrite
\ref{nonreltsallis}-\ref{nonrelnaudts} with all KN means so far considered again generalizing it
to relative information gains, and then replace the so obtained relative R\'{e}nyi entropy in the
exponential expression of \ref{nonrelmasi} (or, proceeding in a somewhat less rigorous manner,
simply extend $\frac{1}{p_{i}} \rightarrow  \frac{p_{i}^{(2)}}{p_{i}^{(1)}}$ in all of them)

\begin{equation}
S_{T} (P,q)= \log_{q} \left<\frac{p_{i}^{(2)}}{p_{i}^{(1)}}\right>_{\!\!\log_{q}} = \frac{1}{1-q}
\, \left[\sum_{i} \, (p_{i}^{(1)})^{q} (p_{i}^{(2)})^{1-q} - 1\right] \, , \label{disdbtsallis}
\end{equation}

\begin{equation}
\hspace{-4mm} S_{R} (P,q)= \log \left<\frac{p_{i}^{(2)}}{p_{i}^{(1)}}\right>_{\!\!\log_{q}} =
\hspace{4mm} \frac{1}{1-q} \, \log \sum_{i} \, (p_{i}^{(1)})^{q} (p_{i}^{(2)})^{1-q} \, ,
\label{disdbrenyi}
\end{equation}

\begin{equation} S_{SM} (P,\{q,r\})=  \log_{r}
\left<\frac{p_{i}^{(2)}}{p_{i}^{(1)}}\right>_{\!\!\log_{q}} = \frac{1}{1-r} \left[ \left( \sum_{i}
(p_{i}^{(1)})^{q} (p_{i}^{(2)})^{1-q} \right) ^{\frac{1-r}{1-q}} - 1 \right] \,
,\label{disdbnaudts}
\end{equation}

\begin{equation}\hspace{4mm} S_{SE}(P,\{q,r\})= \log_{q}
e_{r}^{\,\log \left<\frac{p_{i}^{(2)}}{p_{i}^{(1)}}\right>_{\!\log_{q}}} = \frac{\left[ 1 +
\frac{(1-r)}{(1-q)} \, \log \sum_{i} (p_{i}^{(1)})^{q} (p_{i}^{(2)})^{1-q}
\right]^{\frac{1-q}{1-r}}-1}{1-q}\, .\label{disdbmasi}\end{equation}

\vspace{2mm}

For $p_{i}^{(2)} \rightarrow 1$ they reduce to \ref{tsallis}, \ref{renyi}, \ref{eqit} and
\ref{qrir} respectively, while for $q=1$ Tsallis' and R\'{e}nyi's measures \ref{disdbtsallis} and
\ref{disdbrenyi} become both Kullback's measure \ref{disnegkullback}. From \ref{disdbnaudts}
(\ref{disdbmasi}) we recover R\'{e}nyi's (Tsallis') measure \ref{disdbrenyi} (\ref{disdbtsallis}),
if $r\rightarrow1$, and Tsallis (R\'{e}nyi's) measure \ref{disdbtsallis} (\ref{disdbrenyi}), if
$r\rightarrow q$. Notice how it is much easier to recognize the limits in the
logarithmic-exponential representation.

Straightforwardly  we can now extend to continuous PDFs over parameter spaces $\theta$ and $\phi$.
The \textit{continuous Tsallis, R\'{e}nyi, Sharma-Mittal and supra-extensive relative
information-entropy measures} become

\begin{equation}
\hspace{-50mm}\mathcal{S}_{T} (\mathcal{P},q)= \hspace{0mm} \frac{1}{1-q} \left[ \int \!
p_{1}(\mathbf{x,\theta})^{q} p_{2}(\mathbf{x,\phi})^{1-q} \, d^{n}\mathbf{\!x} -1\right] \, ;
\label{tsallismeanext2}
\end{equation}

\begin{equation}
\hspace{-55mm} \mathcal{S}_{R}(\mathcal{P},q) = \frac{1}{1-q} \log \int \!
p_{1}(\mathbf{x,\theta})^{q} p_{2}(\mathbf{x,\phi})^{1-q} \, d^{n}\mathbf{\!x} \, ;
\label{renyimeanext2}
\end{equation}

\begin{equation} \hspace{-0mm} \mathcal{S}_{SM}(\mathcal{P},\{q,r\}) = \frac{1}{1-r} \left[ \left( \int
p_{1}(\mathbf{x,\theta})^{q} p_{2}(\mathbf{x,\phi})^{1-q} \,d^{n}\mathbf{\!x}
\right)^{\frac{1-r}{1-q}} -1 \right] \, ;\label{naudtsmeanext2}
\end{equation}

\begin{equation} \hspace{-0mm} \mathcal{S}_{SE}(\mathcal{P},\{q,r\}) = \frac{\left[ 1 +
\frac{(1-r)}{(1-q)} \, \log \int (p_{1}(\mathbf{x,\theta}))^{q} (p_{2}(\mathbf{x,\phi}))^{1-q}
\right]^{\frac{1-q}{1-r}}-1}{1-q}\, .\label{masimeanext2}
\end{equation}

Of course, one could again rewrite things all over again, to see that the same result appears if
we extend the Kolmogorov-Nagumo mean to continuity as \begin{equation}\mathcal{S} = f^{-1} \left(
\int \!\! p_{1}(\mathbf{x,\theta}) f(\mathcal{I}_{x}\mathbf{(x,\theta,\phi)}) \, d^{n}\mathbf{\!x}
\right) \, ,\label{contkn}\end{equation} where $$\mathcal{I}_{x}\mathbf{(x,\theta,\phi)}=\log
\left( \frac{p_{2}(\mathbf{x,\phi})}{p_{1}(\mathbf{x,\theta})}\right) \hspace{5mm} \mathrm{for
\,\, extensive \,\, systems} \, ;$$ or
$$\hspace{10mm} \mathcal{I}_{x}\mathbf{(x,\theta,\phi)}=\log_{q} \left(
\frac{p_{2}(\mathbf{x,\phi})}{p_{1}(\mathbf{x,\theta})}\right) \hspace{5mm} \mathrm{for \,\,
non-extensive \,\, systems} \, , $$

and/or using the generalized q-deformed logarithm and exponential expressions from
\ref{nonreltsallis} to \ref{nonrelmasi}, extending $\frac{1}{p_{i}} \rightarrow
\frac{p_{2}(\mathbf{x},\mathbf{\phi})}{p_{1}(\mathbf{x},\mathbf{\theta})} \, .$

Then, applying \ref{contkn} ($f=\log_{q}x$) to obtain the relative and continuous extension of
multiplicity \ref{eii}, one has \footnote{Since we will work with parameters, let us write for a
lighter notation on the multiplicity and the entropies, $\Omega(\mathcal{P},\{q\}) \equiv
\Omega(\theta,\phi)$ and $\mathcal{S}(\mathcal{P},\{q,r\}) \equiv \mathcal{S}(\theta,\phi)$.}
\begin{equation}\Omega(\theta, \phi) =
\left<\frac{p_{2}(x,\phi)}{p_{1}(x,\theta)}\right>_{\log_{q}} = e_{q}^{\int p_{1}(x,\theta)
\log_{q}(\frac{p_{2}(x,\phi)}{p_{1}(x,\theta)})\, d^{n}\mathbf{\!x}} = e_{q}^{S_{T}(\theta, \phi)}
\, . \label{omegacont}\end{equation}

Then we can rewrite \ref{tsallismeanext2} to \ref{masimeanext2} in its relative continuous
extension of \ref{nonreltsallis} to \ref{nonrelmasi} as
\begin{equation}\mathcal{S}_{T}(\theta,\phi)=
\log_{q}\left<\frac{p_{2}(x,\phi)}{p_{1}(x,\theta)}\right>_{\log_{q}} =
\log_{q}\Omega(\theta,\phi) \, ;\label{omegatsallis}\end{equation}

\begin{equation}\mathcal{S}_{R}(\theta,\phi)=
\log\left<\frac{p_{2}(x,\phi)}{p_{1}(x,\theta)}\right>_{\log_{q}} = \log\Omega(\theta,\phi) \, ;
\label{omegarenyi}\end{equation}

\begin{equation}\mathcal{S}_{SM}(\theta,\phi)=
\log_{r}\left<\frac{p_{2}(x,\phi)}{p_{1}(x,\theta)}\right>_{\log_{q}} =
\log_{r}\Omega(\theta,\phi) \, ; \label{omegasm}\end{equation}

\begin{equation}\mathcal{S}_{SE}(\theta,\phi)= \log_{q}e_{r}^{\log
\left<\frac{p_{2}(x,\phi)}{p_{1}(x,\theta)}\right>_{\log_{q}}} =
\log_{q}e_{r}^{\log\Omega(\theta,\phi)} \, . \label{omegase}\end{equation}

\section{The role of Fisher information for generalized entropy measures}

\subsection{The Fisher information measure}

We are now ready to proceed towards the real aim of this paper. We begin with a brief introduction
to Fisher information.

In 1921, R. Arnold Fisher defined an information measure which could account for the "quality" or
"efficiency" of a measurement. Calling \textit{efficient estimator} or \textit{best estimator},
the best unbiased estimate $\widehat{\theta}(x)$ of $\theta$ after many independent measurements
on a random variable $x$ such that $<\!\widehat{\theta}(x)\!\!> \, = \, \int p(x,\theta) \,
\widehat{\theta}(x) \, dx = \theta$, Fisher defined the \textit{efficiency} or \textit{quality} of
a measurement, $\mathcal{I}_{F}$, the quantity which satisfies $$\mathcal{I}_{F} \, e^{2}= 1 \,
,$$ where $e^{2}=\int p(x,\theta) \, [\widehat{\theta}(x) - \theta]^{2} \, dx$ is the mean square
error.

Fisher showed \cite{Fisher1} that then $\mathcal{I}_{F}$ is uniquely identified as
$$\mathcal{I}_{F}(\mathcal{P}) = \left< \left( \frac{\partial \log
p(x,\theta)}{\partial \theta} \right) ^{2} \right>_{lin} = $$ $$ = \int_{\mathcal{S}_{sp}}
p(x,\theta) \left( \frac{\partial \log p(x,\theta)}{\partial \theta} \right)^{2} \, dx \, =
\int_{\mathcal{S}_{sp}} \frac{1}{p(x,\theta)} \left( \frac{\partial p(x,\theta)}{\partial \theta}
\right)^{2} \, dx \, .$$

For any other estimator one chooses, the \textit{Cramer-Rao inequality}, or \textit{Cramer-Rao
bound}, holds $$\mathcal{I}_{F} \, e^{2} \geq 1 \, .$$

Going over to N-dimensional vector random variables $\mathbf{x} =(x_{1},...,x_{N})$ on an
M-dimensional parameter space $\mathbf{\theta} = (\theta_{1},...,\theta_{M})$, Fisher defined its
celebrated (symmetric) \textit{Fisher information matrix} which elements are given by
\begin{equation} \hspace{-28mm} F_{ij}(\mathbf{\theta}) =  \left<  \frac{\partial \log
p(\mathbf{x,\theta})}{\partial \theta_{i}} \, \frac{\partial \log p(\mathbf{x,\theta})}{\partial
\theta_{j}} \right>_{lin} $$ $$= \int_{\mathcal{S}_{sp}} p(\mathbf{x,\theta}) \, \frac{\partial
\log p(\mathbf{x,\theta})}{\partial \theta_{i}} \, \frac{\partial \log
p(\mathbf{x,\theta})}{\partial \theta_{j}} \, d^{n}\mathbf{\!x} $$ $$\hspace{-11mm}=
\int_{\mathcal{S}_{sp}} \frac{1}{p(\mathbf{x,\theta})} \, \frac{\partial
p(\mathbf{x,\theta})}{\partial \theta_{i}} \, \frac{\partial p(\mathbf{x,\theta})}{\partial
\theta_{j}} \, d^{n}\mathbf{\!x} \, , \label{fishermatrix}
\end{equation}

with $(i,j=1,..,M)$. If we would further extend to an L-dimensional continuous probability space
$\mathcal{P}=(p_{1},...,p_{L})$, then the most general expression for Fisher information writes
$$\mathcal{I}_{F}(\mathcal{P}) = \sum_{k=1}^{L} \, \sum_{i,j=1}^{M}
F_{ij}^{k}(\mathbf{\theta}) \, .$$

\subsection{The Fisher information matrix as a metric tensor}

We will not go into the details in what would be a much too long exposition of information
geometry and shall highlight only in an introductory manner the status of the FIM as a metric
tensor for a statistical manifold (for a more rigorous account of the subject see e.g.
\cite{amari}, \cite{Corcuera}, \cite{Wagenaar}, \cite{Rodriguez}, and references therein).

Consider a family of $\mathcal{C}^{2}$ differentiable PDFs with N-dimensional continuous vector
random variables $\mathbf{x}$, parametrized by an M-dimensional continuous real vector parameter
space $\mathbf{\theta}$ on an open interval $I_{\mathbf{\theta}} \subseteq \mathbb{R}^{M}$
$$\mathcal{F}_{\theta} = \{p(\mathbf{x,\theta}) \in \mathcal{C}^{2}; \theta \in
I_{\mathbf{\theta}}\} \, .$$ The notion of a \textit{differential statistical manifold} is
identified in the fact that the parameters $\theta$ can be conceived as providing a local
coordinate system for an M-dimensional manifold $\mathcal{M}$ which points are in a one to one
correspondence with the distributions $p \in \mathcal{F}_{\theta}$.

Since information-entropy measures are log-probability functionals defined on $\mathcal{M}$, it is
convenient to consider also the function $l: \mathcal{M} \rightarrow \mathbb{R}$ on the manifold
$\mathcal{M}$ defined as $l(\mathbf{\theta}) \equiv \log p(\mathbf{x,\theta})$. This is commonly
called the \textit{log-likelihood function}. Labelling the manifold's tangent space
$T_{\theta}(\mathcal{M})$, the directional derivatives of $l(\theta)$ along the tangent vectors
$\widehat{\mathbf{e}}_{i} \in T_{\theta}(\mathcal{M})$ at a point in $\mathcal{M}$ with
coordinates $\mathbf{\theta}$ are (use the shorthand $\partial_{i} \equiv \frac{\partial}{\partial
\theta_{i}}$):  $\,\,\partial_{i}l(\mathbf{\theta}) \, \widehat{\mathbf{e}}_{i} =
\frac{\partial_{i}p(\mathbf{x,\theta})}{p(\mathbf{x,\theta})}  \, \widehat{\mathbf{e}}_{i} \, .$

FIM \ref{fishermatrix} can also be seen as the expectation value with respect to
$p(\mathbf{x,\theta})$ of the partial derivatives of $l(\mathbf{\theta})$, which is the reason why
in the literature it is frequently written as $$F_{ij}(\mathbf{\theta})= E[\partial_{i}
l(\mathbf{\theta}) \, \partial_{\!j} l(\mathbf{\theta})] \, .$$ This is a symmetric,
non-degenerate, bilinear form on a vector space of random variables $\partial_{i}
l(\mathbf{\theta})$. But a Riemannian metric g is per definition a symmetric non-degenerate inner
product on the manifold's tangent space $T_{\theta}(\mathcal{M})$, and one can therefore consider
the FIM as the statistical analogue of the metric tensor for a statistical manifold.

By the way, it is worth mentioning that Corcuera \& Giummol`e showed \cite{Corcuera} that the FIM
has also the unique properties of being covariant under reparametrization of the parameter space
of the manifold, and invariant under reparametrization of the sample space (see also Wagenaar
\cite{Wagenaar} for a review). This is an appealing aspect which possibly suggests that Fisher
information might play some role in future quantum spacetime theories.

Now, the metric tensor tells how to compute the distance between any two points in a given space.
Here we are considering the distance between two points on a statistical differential manifold
mapped on a measure functional, i.e. the informational difference between them. This idea can be
introduced regarding Kullback's relative information measure to account for the net dissimilarity
between two families of PDFs with parameters, $\theta$ and $\phi$. Intuitively one can imagine
this as measuring a "distance" between these two families. However, strictly speaking, this is not
a metric distance because it is neither symmetric nor satisfies the triangle inequality (on
statistical manifolds one has to consider an extended version of Pythagora's law). The symmetry
condition however can be restored if instead of the single information measure we use the
\textit{divergence} $\mathcal{D}$ of two PDFs, $p_{1}$ and $p_{2}$, defined as\footnote{The notion
of divergence in information geometry can be established in a rigorous way and is much more
general. We shall however use only this particular type of definition because it is sufficient for
our purposes.} $$\mathcal{D}(p_{1},p_{2}) = \frac{\mathcal{S}(p_{1},p_{2}) +
\mathcal{S}(p_{2},p_{1})}{2} \, .$$ If we choose to set \begin{equation} p_{1}(\mathbf{x,\theta})
\equiv p \, (\mathbf{x},\theta) ; \,\,\, \ p_{2}(\mathbf{x,\phi}) \equiv p\,(\mathbf{x},\theta
+d\theta)\, , \label{p1p2}\end{equation} then the symmetric divergence
$\mathcal{D}(p\,(\mathbf{x},\theta),p\,(\mathbf{x},\theta +d\theta)) \equiv \mathcal{D}(\theta,
\theta +d\theta)$ can be intended as an extension of the square of the Riemannian distance between
two nearby distributions. Expanded to second order it gives $$\mathcal{D}(\theta, \theta +d\theta)
= \frac{1}{2!} \sum_{ij} \left[ \frac{\partial^{2} \mathcal{D} (\theta,\phi)}{\partial \theta_{i}
\partial \theta_{j}} \right]_{\phi=\theta}\!\!\!\!\!\!\!\!d\theta_{i}d\theta_{j} \,+\, O(d
\theta^{3}) \, ,$$ because $\mathcal{D}(\theta, \phi)$ is minimal at $\phi=\theta$ and the first
order vanishes. It is the second order, not the first, which is the leading one in every
information measure divergence, and it can be shown (\cite{amari}, \cite{Corcuera},
\cite{Wagenaar}) that it is the second derivative of the divergence which defines the metric, i.e.
\begin{equation} g_{ij}(\mathbf{\theta}) = \left[ \frac{\partial^{2} \mathcal{D}
(\theta,\phi)}{\partial \theta_{i} \partial \theta_{j}} \right] _{\phi=\theta} = \frac{1}{2} \,
\left[ \frac{\partial^{2} \! \left( \mathcal{S} (\theta,\phi) + \mathcal{S}(\phi,\theta)
\right)}{\partial \theta_{i}\partial \theta_{j}}\right] _{\phi=\theta} \, .
\label{metrictensor2}\end{equation}

In case of Kullback's measure \ref{negkullback}, the divergence is defined as
$$\mathcal{D}_{KL}(\theta, \phi) = \frac{1}{2} \int \left[p(\mathbf{x,\theta}) -
p(\mathbf{x,\phi}) \right] \, \log \frac{p(\mathbf{x,\phi})}{p(\mathbf{x,\theta})} \,
d^{n}\mathbf{\!x} \, .$$ From \ref{metrictensor2}, and keeping in mind that if we want the
normalization condition to hold for every $\theta$ implies \begin{equation}
\frac{\partial}{\partial \theta_{i}}\, 1 = \frac{\partial}{\partial \theta_{i}} \int
p(\mathbf{x,\theta}) d^{n}\mathbf{\!x} = \int \frac{\partial}{\partial \theta_{i}}
p(\mathbf{x,\theta}) d^{n}\mathbf{\!x} = \int \frac{\partial^{2}}{\partial \theta_{i} \theta_{j}}
p(\mathbf{x,\theta}) d^{n}\mathbf{\!x} = 0 \, , \label{normalization}\end{equation} we have
$$g_{ij}^{KL}(\mathbf{\theta}) = - \int \frac{1}{p(\mathbf{x,\theta})} \, \frac{\partial
p(\mathbf{x,\theta})}{\partial \theta_{i}} \, \frac{\partial p(\mathbf{x,\theta})}{\partial
\theta_{j}} \, d^{n}\mathbf{\!x} = - F_{ij}(\mathbf{\theta}) \, ,$$ which is the (i,j)-th element
of the negative FIM \ref{fishermatrix}.

This is a very important and known result from information geometry. It is in this sense that
$g_{ij}$ can be seen as a metric tensor which measures a "distance" on a statistical manifold in a
Riemannian space. In this sense Fisher information can be said to be a sort of "mother information
measure".

\section{The Fisher metric for generalized information-entropy measures}

We can generalize this result of information geometry. The Fisher metric for Tsallis, R\'{e}nyi,
the Sharma-Mittal and the supra-extensive measures can be obtained considering the relative
entropy measures as defined in \ref{tsallismeanext2}, \ref{renyimeanext2}, \ref{naudtsmeanext2}
and \ref{masimeanext2} respectively (with $p_{1}=p(\mathbf{x,\theta})$,
$p_{2}=p(\mathbf{x,\phi})$), from their respective symmetric \textit{divergence}
$$\mathcal{D}(\theta,\phi) = \frac{\mathcal{S}(\theta,\phi) + \mathcal{S}(\phi,\theta)}{2} \, ,$$
defined on $\mathcal{F}_{\theta}$.

What we need is the evaluation of \ref{metrictensor2} for each information measure
\ref{tsallismeanext2} to \ref{masimeanext2}. One can of course compute directly the (somewhat
fuzzy) second derivatives $\frac{\partial^{2} \mathcal{S}(\phi,\theta)}{\partial \theta_{i}
\partial \theta_{i}}$ each time (and for each $\theta \rightarrow \phi$ parameter exchange).
However, the q-deformed generalized logarithm and exponential formalism and the KN-representation
make this task easier since it needs only the evaluation of Tsallis' entropy, the rest follows
almost automatically.

The final result will be that $g_{ij}^{KL}$ remains still the fundamental quantity, but for these
more general (supra-extensive, Sharma-Mittal, R\'{e}nyi and Tsallis) relative entropies the
statistical metric tensor ($g^{SE}_{ij}, g^{SM}_{ij}, g^{R}_{ij}$ and $g^{T}_{ij}$ respectively)
turns out to be only slightly extended by a scalar multiplicative q-deforming factor as
\begin{equation}g^{SE}_{ij}(\mathbf{\theta}) = g^{SM}_{ij}(\mathbf{\theta}) =
g^{R}_{ij}(\mathbf{\theta}) = g^{T}_{ij}(\mathbf{\theta}) = q \, g^{KL}_{ij}(\mathbf{\theta}) = -q
\, F_{ij}(\mathbf{\theta}) \, .\label{allgij} \end{equation}

This shows also that while $g_{ij}$ depends from the q-deforming parameter it is independent from
the r-extensivity parameter. This is quite natural since Fisher information accounts for the
"quality" of a measure, or so to say, our "differential capacity to distinguish" locally between
two neighboring PDFs, and this in turn depends from the "form" of the PDF (the q-scaling), but is
independent from the extensive, non-extensive or supra-extensive character, since these are global
features of the system. We shall see how it is the normalization condition imposed on PDFs that
leads to this independency (and recover the known fact that this is also the same reason why
$g_{ij}$ is symmetric). Moreover, it will also become clear how Fisher information measures the
rate of change of the multiplicity under a parameter variation.

\subsection{Fisher from Tsallis information}

First of all consider the derivation rules \begin{equation}\frac{\partial \log_{q}x}{\partial x} =
\frac{1}{x^{q}} \, ; \hspace{8mm} \frac{\partial e_{q}^{x}}{\partial x}=
\left(e_{q}^{x}\right)^{q} \, .\label{deriv}
\end{equation}

Writing Tsallis' continuous relative entropy \ref{tsallismeanext2} in the q-deformed Shannon
notation of \ref{q-shannon}, we have \begin{equation}\mathcal{S}_{T}(\theta, \phi) = \int
p_{1}(\mathbf{x,\theta}) \log_{q} \left( \frac{p_{2}(\mathbf{x,\phi})}{p_{1}(\mathbf{x,\theta})}
\right) \, d^{n}\mathbf{\!x} \, .\label{secondtsallis}\end{equation} We must be careful in
remembering that in general the entropy measures considered are not symmetric and have to consider
also $$\mathcal{S}_{T}(\mathbf{\phi},\mathbf{\theta}) = \int p_{2}(\mathbf{x,\phi}) \log_{q}
\left( \frac{p_{1}(\mathbf{x,\theta})}{p_{2}(\mathbf{x,\phi})} \right) \, d^{n}\mathbf{\!x} \, .$$

Then, applying the q-logarithm derivation rule \ref{deriv}, one obtains for the first case (as
before $\frac{\partial}{\partial \theta_{i}} \equiv \partial_{i}; \, \frac{\partial^{2}}{\partial
\theta_{i} \partial \theta_{j}} \equiv \partial_{ij}$)
\begin{equation} \partial_{i}\mathcal{S}_{T}(\mathbf{\theta},\mathbf{\phi}) = \int \left[ \log_{q}
\left( \frac{p_{2}(\mathbf{x,\phi})}{p_{1}(\mathbf{x,\theta})}\right) - \left(
\frac{p_{2}(\mathbf{x,\phi})}{p_{1}(\mathbf{x,\theta})}\right)^{1-q} \right]
\partial_{i}p_{1}(\mathbf{x,\theta}) \, d^{n}\mathbf{\!x} \, ,\label{tsallisderive}\end{equation}
and $$\partial_{ij}\mathcal{S}_{T}(\mathbf{\theta},\mathbf{\phi}) = - \, q \int \left(
\frac{p_{2}(\mathbf{x,\phi})}{p_{1}(\mathbf{x,\theta})} \right)^{1-q} \!\!\!\!\!
\frac{1}{p_{1}(\mathbf{x,\theta})} \,\, \partial_{i} p_{1}(\mathbf{x,\theta}) \partial_{j}
p_{1}(\mathbf{x,\theta})  \, d^{n}\mathbf{\!x} \, $$ $$\hspace{36mm} + \,  \int \left[ \log_{q}
\left( \frac{p_{2}(\mathbf{x,\phi})}{p_{1}(\mathbf{x,\theta})}\right) - \left(
\frac{p_{2}(\mathbf{x,\phi})}{p_{1}(\mathbf{x,\theta})}\right)^{1-q} \right] \,
\partial_{ij}p_{1}(\mathbf{x,\theta}) d^{n}\mathbf{\!x} \, .$$

While in the second case one has quite different derivatives
\begin{equation}\partial_{i}\mathcal{S}_{T}(\mathbf{\phi},\mathbf{\theta}) = \int \left(
\frac{p_{2}(\mathbf{x,\phi})}{p_{1}(\mathbf{x,\theta})} \right)^{q}
\partial_{i}p_{1}(\mathbf{x,\theta}) \, d^{n}\mathbf{\!x} \, ,
\label{tsallisderive2}\end{equation} and
$$\partial_{ij}\mathcal{S}_{T}(\mathbf{\phi},\mathbf{\theta})= - q \int \left(
\frac{p_{2}(\mathbf{x,\phi})}{p_{1}(\mathbf{x,\theta})} \right)^{q}
\frac{1}{p_{1}(\mathbf{x,\theta})} \,\, \partial_{i} p_{1}(\mathbf{x,\theta}) \partial_{j}
p_{1}(\mathbf{x,\theta})  \, d^{n}\mathbf{\!x} \, +$$ $$\hspace{-8mm}+ \int \left(
\frac{p_{2}(\mathbf{x,\phi})}{p_{1}(\mathbf{x,\theta})} \right)^{q}
\partial_{ij}p_{1}(\mathbf{x,\theta}) \, d^{n}\mathbf{\!x} \, .$$

Note that these derivatives are not the same that one would obtain directly from
\ref{tsallismeanext2}, because in that case one assumes implicitly the normalization condition
satisfied a priori. \ref{tsallismeanext2} and \ref{secondtsallis} are numerically identical only
for a normalized PDF. The logarithmic-exponential representation, as in the latter case, does
therefore not only represent a more general expression but, highlights better where and with what
effects the normalization enters into the play. Restricting to PDFs as \ref{p1p2}, then, because
of the normalization condition \ref{normalization}, from \ref{tsallisderive} and
\ref{tsallisderive2} one obtains
\begin{equation}\left[\partial_{i}\mathcal{S}_{T}(\mathbf{\theta},\mathbf{\phi})\right]_{\phi=\theta}
= \left[\partial_{i}\mathcal{S}_{T}(\mathbf{\phi},\mathbf{\theta})\right]_{\phi=\theta} =0 \,
,\label{tpderive}\end{equation}

while remembering the expression for the FIM \ref{fishermatrix}
$$\hspace{-5mm} \left[\partial_{ij}\,\mathcal{S}_{T}(\mathbf{\theta},\mathbf{\phi})\right]_{\phi=\theta}=
\left[\partial_{ij}\,\mathcal{S}_{T}(\mathbf{\phi},\mathbf{\theta})\right]_{\phi=\theta} =$$
$$\hspace{20mm}= - \, q \! \int \frac{1}{p(\mathbf{x,\theta})} \,\, \partial_{i}
p(\mathbf{x,\theta}) \partial_{j} p(\mathbf{x,\theta})  \, d^{n}\mathbf{\!x}  = -q F_{ij}(\theta)
\, ,$$ which, through \ref{metrictensor2}, gives us finally $g_{ij}^{T} = - q
F_{ij}(\mathbf{\theta})$.

So, since the FIM is symmetric, by the way, we see that in this case, and as we shall see also in
all the others, it is in particular the normalization condition which renders the statistical
metric tensor $g_{ij}$ symmetric.

\subsection{Fisher from R\'{e}nyi information}

Evaluating Tsallis' derivatives is indispensable but, once established, we don't need to make any
direct derivative anymore for all the other measures if we work with generalized logarithms and
exponentials. We don't even need to repeat the derivation for the symmetry considerations.

In fact, \ref{metrictensor2} for R\'{e}nyi's measure can be obtained from \ref{ireqt}. From
\ref{deriv} we obtain (the arguments $(\mathbf{x,\theta})$ or $(\mathbf{x,\phi})$ of the measures,
the PDFs or of the FIM, shall be omitted if it is not needed otherwise)
\begin{equation}\partial_{i}\mathcal{S}_{R}=
\partial_{i} \log e_{q}^{\mathcal{S}_{T}} = \left(e_{q}^{\mathcal{S}_{T}}\right)^{q-1}
\partial_{i}\mathcal{S}_{T} \, ,\label{renyiderive}\end{equation} and
$$\partial_{ij}\mathcal{S}_{R}= \left(e_{q}^{\mathcal{S}_{T}}\right)^{q-1} \left[
\partial_{ij}\mathcal{S}_{T} + (q-1) \left(e_{q}^{\mathcal{S}_{T}}\right)^{q-1}
\partial_{i}\mathcal{S}_{T} \partial_{j}\mathcal{S}_{T} \right] \, .$$

Since $[\mathcal{S}_{T}]_{\phi=\theta}= 0$, applying the normalization condition (i.e. because of
\ref{tpderive}), we have \begin{equation}\left[\partial_{ij}\mathcal{S}_{R}\right]_{\phi=\theta} =
\left[\partial_{ij}\mathcal{S}_{T} \right]_{\phi=\theta} \, ,\label{tsarenyi}\end{equation}

which leads us to state $g_{ij}^{R}=g_{ij}^{T}=-qF_{ij}$.

\subsection{Fisher from Sharma-Mittal information}

Use Sharma-Mittal entropy as given in \ref{eqit} and proceed as in the previous case
$$\partial_{i}\mathcal{S}_{SM} = \partial_{i} \log_{r} e_{q}^{\mathcal{S}_{T}} =
\left(e_{q}^{\mathcal{S}_{T}}\right)^{q-r} \partial_{i}\mathcal{S}_{T} \, ,$$ and $$
\partial_{ij}\mathcal{S}_{SM} = \left(e_{q}^{\mathcal{S}_{T}}\right)^{q-r}
\left[\partial_{ij}\mathcal{S}_{T} + (q-r) \left(e_{q}^{\mathcal{S}_{T}}\right)^{q-1}
\partial_{i}\mathcal{S}_{T} \, \partial_{j} \mathcal{S}_{T} \right] \, .$$

\vspace{2mm}

And again because of \ref{tpderive} $$\left[\partial_{ij}\mathcal{S}_{SM}\right]_{\phi=\theta} =
\left[\partial_{ij}\mathcal{S}_{T}\right]_{\phi=\theta} \, ,$$ we have again
$g_{ij}^{SM}=g_{ij}^{T}=-qF_{ij}$. Note that it is the normalization condition, forcing the r.h.s.
derivatives to vanish, which leads to the independency of $g_{ij}$ from the non-extensivity
parameter $r$.

\subsection{Fisher from supra-extensive information}

From \ref{qrir} we get $$\partial_{i} \mathcal{S}_{SE} = \partial_{i}
\log_{q}e_{r}^{\mathcal{S}_{R}} = \left(e_{r}^{\mathcal{S}_{R}}\right)^{r-q}
\partial_{i}\mathcal{S}_{R} \, ,$$ and $$\partial_{ij} \mathcal{S}_{SE} =
\left(e_{r}^{\mathcal{S}_{R}}\right)^{r-q} \left[ \partial_{ij}\mathcal{S}_{R} + (r-q)\
\left(e_{r}^{\mathcal{S}_{R}}\right)^{r-1} \partial_{i}\mathcal{S}_{R} \,
\partial_{j}\mathcal{S}_{R} \right] \, .$$

Because of \ref{tpderive} and \ref{renyiderive}
$$\left[\partial_{i}\mathcal{S}_{R}(\theta, \phi)\right]_{\phi=\theta} =
\left[\partial_{i}\mathcal{S}_{R}(\phi, \theta)\right]_{\phi=\theta} =0 \, ,$$ then, remembering
\ref{tsarenyi} one has $\left[\partial_{ij}\mathcal{S}_{SE}\right]_{\phi=\theta} =
\left[\partial_{ij}\mathcal{S}_{R}\right]_{\phi=\theta} = \left[ \partial_{ij}\mathcal{S}_{T}
\right]_{\phi=\theta}$, and finally $$g_{ij}^{SE}=g_{ij}^{T} = -q F_{ij} \, .$$

Therefore, either $\partial_{ij}\mathcal{S}_{SM}$ as $\partial_{ij}\mathcal{S}_{SE}$ don't depend
from the $r$ parameter because of the normalization condition.

\subsection{Working with the multiplicity}

Just for didactics, in order to show how the generalized exponential-logarithmic formalism
combined with the KN expressions can be used, we reach the same conclusion from the perspective of
the entropies as a measure of multiplicity. From \ref{omegacont} one has
$$\partial_{i}\Omega = \left(e_{q}^{S_{T}}\right)^{q} \partial_{i}S_{T} = \Omega^{\,q} \,
\partial_{i}S_{T} \,\,\, \Rightarrow \,\, \left[\partial_{i}\Omega(\theta,\phi)\right]_{\phi=\theta} =
\left[\partial_{i}\Omega(\phi,\theta)\right]_{\phi=\theta} = 0 \, ,$$ while
$$\partial_{ij}\Omega = \Omega^{\,q} \!\left[\partial_{ij}S_{T} + q \, \Omega^{\,q-1}
\partial_{i} S_{T} \partial_{j}S_{T} \right] \, ,$$ which implies that
$$\left[\partial_{ij}\Omega\right]_{\phi=\theta} = \left[\partial_{ij}S_{T}\right]_{\phi=\theta} = -q F_{ij} \, .$$

Therefore, working with information-entropy measures expressed with the multiplicity as in
\ref{omegatsallis} to \ref{omegase}, the SM's measure second derivative is
$$\partial_{ij}S_{SM}= \frac{\partial_{ij}\Omega}{\Omega^{\,r}}- r \frac{\partial_{i}\Omega \,
\partial_{j}\Omega}{\Omega^{r+1}} \, ,$$ one has $$\left[\partial_{ij}S_{SM}\right]_{\phi=\theta} =
- q F_{ij} \, ,$$ and the above results for Tsallis, R\'{e}nyi and Shannon's measure all follow
again as special cases.

Finally, for SE measure $$\partial_{ij}S_{SE}=\partial_{j}\left(\left( e_{r}^{\log
\Omega}\right)^{r-q} \frac{\partial_{i}\Omega}{\Omega}\right)$$ $$=(r-q) \left(e_{r}^{\log
\Omega}\right)^{2r-q-1} \, \frac{\partial_{i}\Omega}{\Omega}$$ $$+ \left(e_{r}^{\log
\Omega}\right)^{r-q} \, \left( \frac{\partial_{ij}\Omega}{\Omega} -
\frac{\partial_{i}\Omega\,\partial_{j}\Omega}{\Omega^{2}} \right) \, ,$$

and, as was to expect, the final results simplifies to
$$\left[\partial_{ij}S_{SE}\right]_{\phi=\theta} = [\partial_{ij}\Omega]_{\phi=\theta}= - q F_{ij} \, .$$

Therefore, since the second order of the multiplicity is the leading one, we can say that Fisher
information accounts (times a negative parameter multiplicative deformation factor) for the change
of multiplicity (the change of number of microstates of a system) under a statistical parameter
variation. This is another way to interpret the fundamental connection between Fisher information
and entropy measures.

\section{Conclusion}

Using the notion of Kullback-Leibler's relative entropy, generalizing it to all entropies, we
showed, as it was already known for Kullback's measure, that once again the FIM appears as the
same statistical metric tensor \ref{allgij} for Tsallis, R\'{e}nyi, Sharma-Mittal and the
supra-extensive measures too. The differential-geometric properties of the divergence for each
measure are independent from the extensive, non-extensive or supra-extensive character of the
system, but depend only from the q-deforming parameter. This independency and the symmetry of
$g_{ij}$ are guaranteed by the normalization condition. We could also see how Fisher information
has to be interpreted as a quantity proportional to the change of the information multiplicity
under the statistical parameter variation. Generally, the derivation of Fisher information proved
to be easier to obtain by exploiting the q-deformed logarithm and exponential formalism or the
KN-representation of information-entropy measures. The overall global picture of the
generalization process we have undertaken so far can be finally summarized in the diagram of the
following page (where $p_{1}$ and $p_{2}$ can be both PDs or PDFs).

\newpage

\setlength{\unitlength}{1mm}

\begin{picture}(30,95)(8,-3)

\put(56,59){\framebox(56,18)}

\put(30,87){\Large \textit{Hierarchy of generalized relative entropy measures} \normalsize}

\put(70,72){\Large Fisher Matrix \normalsize}

\put(62,64){\Large \shortstack{$-q \left< \frac{\partial \log p}{\partial \theta_{i}} \,
\frac{\partial \log p}{\partial \theta_{j}} \right>_{\!\mathrm{lin}}$} \normalsize}

\put(85,54){\vector(0,1){5}}

\put(78,51.5){\normalsize $p_{2} \rightarrow p_{1} \equiv p$ \normalsize}

\put(85,45){\line(0,1){5}}

\put(0,-115){\framebox(170,160)}

\put(10,32){\large Sharma-Mittal  \normalsize}

\put(14,20){\large$\log_{r} \left<\frac{p_{2}}{p_{1}}\right>_{\!\!\log_{q}}$ \normalsize}

\put(26,16){\line(1,-1){10}}

\put(35,3){\normalsize $r \rightarrow 1$ \normalsize}

\put(41,1){\vector(1,-1){10}}

\put(43,21){\line(1,0){10}}

\put(53,21){\line(2,-1){10}}

\put(63,13.5){\normalsize $r \rightarrow q$ \normalsize}

\put(71,12){\vector(2,-1){40}}

\put(120,32){\large Supra-extensive  \normalsize}

\put(127,20){\large$\log_{q} \! e_{r}^{\log \left<\frac{p_{2}}{p_{1}}\right>_{\!\!\log_{q}}}$
\normalsize}

\put(140,16){\line(-1,-1){10}}

\put(125,3){\normalsize $r \rightarrow 1$ \normalsize}

\put(125,1){\vector(-1,-1){10}}

\put(122,21){\line(-1,0){12}}

\put(110,21){\line(-2,-1){10}}

\put(92,13.5){\normalsize $r \rightarrow q$ \normalsize}

\put(95,12){\vector(-2,-1){40}}

\put(46,-14){\large R\'{e}nyi \normalsize}

\put(43,-22){\large \shortstack{$\log \left<\frac{p_{2}}{p_{1}}\right>_{\!\!\log_{q}}$}
\normalsize}

\put(53,-28){\line(1,-1){10}}

\put(61,-41.5){\normalsize $q \rightarrow 1$ \normalsize}

\put(68,-43){\vector(1,-1){10}}

\put(109,-14){\large Tsallis \normalsize}

\put(105,-22){\large \shortstack{$\log_{q} \left<\frac{p_{2}}{p_{1}}\right>_{\!\!\log_{q}}$}
\normalsize}

\put(114,-28){\line(-1,-1){10}}

\put(97,-41.5){\normalsize $q \rightarrow 1$ \normalsize}

\put(100,-43){\vector(-1,-1){10}}

\put(67,-59){\large Kullback-Leibler \normalsize}

\put(74,-67){\large \shortstack{$\log \left<\frac{p_{2}}{p_{1}}\right>_{\!\!\log}$} \normalsize}

\put(84,-71){\line(0,-1){5}}

\put(78.5,-80){\normalsize $p_{2} \rightarrow 1$ \normalsize}

\put(84,-82){\vector(0,-1){5}}

\put(56,-93){\large Shannon (Boltzmann-Gibbs) \normalsize}

\put(75,-101){\large \shortstack{$\log \left<\frac{1}{p_{1}}\right>_{\!\!\log}$} \normalsize}

\end{picture}

\end{document}